\documentclass[12pt]{article}
\usepackage{epsf}
\usepackage{bm,physics}
\usepackage{cite}
\usepackage{color}
\usepackage{amsmath,amssymb}
\usepackage{graphicx}
\usepackage[colorlinks,citecolor=blue]{hyperref}
\usepackage{comment}
\let\originalleft\left
\let\originalright\right
\renewcommand{\left}{\mathopen{}\mathclose\bgroup\originalleft}
\renewcommand{\right}{\aftergroup\egroup\originalright}

\renewcommand{\thefootnote}{\fnsymbol{footnote}}
\def\thefootnote{\fnsymbol{footnote}}

\makeatletter
\renewcommand{\theequation}{%
\thesection.\arabic{equation}}
\@addtoreset{equation}{section}
\makeatother

\begin{document}

\begin{titlepage}

\begin{center}

\vskip .45in

\bigskip\bigskip
{\Large \bf EFT of Dark Energy with Cosmic Chronometers: Reconstructing Background EFT Functions}

\vskip .65in

{\large 
Fumiya~Okamatsu$^{1}$ and
Kazufumi~Takahashi$^{1,2}$
\vspace{2mm} \\
}
\vskip 0.2in

{\em 
$^{1}$Department of Physics, College of Humanities and Sciences, Nihon University, Tokyo 156-8550, Japan\\
$^{2}$Center for Gravitational Physics and Quantum Information, Yukawa Institute for Theoretical Physics, Kyoto University, 606-8502, Kyoto, Japan
}

\end{center}
\vskip .5in

\begin{abstract}
The effective field theory (EFT) of dark energy provides a model-independent framework for studying cosmology within scalar-tensor theories. In this work, we explore how the time evolution of the cosmological background, inferred from cosmic chronometer measurements of the Hubble parameter, can be used to reconstruct the relevant EFT functions. Our approach enables the direct determination of these EFT functions from observational data without assuming any specific cosmological model. This makes it possible to test the background evolution of a wide range of dark energy models, including the $\Lambda$CDM (cold dark matter) model. We further demonstrate how the reconstructed EFT functions can be applied to constrain concrete theories, such as the quintessence model.
\end{abstract}

\end{titlepage}

\renewcommand{\thepage}{\arabic{page}}
\setcounter{page}{1}
\renewcommand{\thefootnote}{\#\arabic{footnote}}
\setcounter{footnote}{0}

\section{Introduction} \label{sec:intro}

The $\Lambda$CDM (cold dark matter) model, based on general relativity, has been highly successful in explaining a wide range of observations---including the accelerated expansion of the Universe, the cosmic microwave background~(CMB), and the power spectrum and statistical properties of large-scale structure---and has consequently become established as the standard model of cosmology. Meanwhile, there is a significant discrepancy of about $5\sigma$ between the direct measurement of the Hubble constant~$H_0$ by the SH$0$ES collaboration~\cite{Riess:2021jrx} and the value inferred by the Planck collaboration from CMB observations~\cite{Planck:2018vyg}. This so-called ``Hubble tension'' may be a sign of new physics, and a variety of theoretical explanations have been proposed~\cite{Verde:2019ivm}, including modifications of gravity~\cite{Renk:2017rzu, Kreisch:2019yzn, Braglia:2020auw}. Even apart from this context, modified gravity has been studied extensively as a framework for comparison with general relativity~\cite{Koyama:2015vza,Ferreira:2019xrr,Arai:2022ilw}.

Modified gravity models involve additional degree(s) of freedom in general. Among them, the class of scalar-tensor theories, involving a single scalar degree of freedom on top of the spacetime metric, provides a simple and robust framework. The most general class of scalar-tensor theories with second-order Euler-Lagrange equations is known as the Horndeski theory~\cite{Horndeski:1974wa,Deffayet:2011gz,Kobayashi:2011nu}, which accommodates, e.g., $f(R)$ gravity, quintessence, and k-essence models. Moreover, various extensions of the Horndeski theory have been developed recently~\cite{Langlois:2015cwa,Crisostomi:2016czh,BenAchour:2016fzp,DeFelice:2018mkq,DeFelice:2021hps,Takahashi:2021ttd,Takahashi:2022mew,Takahashi:2023jro,Takahashi:2023vva} (see also Refs.~\cite{Langlois:2018dxi,Kobayashi:2019hrl} for reviews).

When studying cosmology within such scalar-tensor theories, the so-called effective field theory~(EFT) of inflation/dark energy provides a universal description of perturbations about the Friedmann-Lema{\^i}tre-Robertson-Walker~(FLRW) background~\cite{Creminelli:2006xe,Cheung:2007st,Gubitosi:2012hu}. The only assumption of the EFT is that the background scalar field has a timelike gradient; therefore, it can in principle be applied to arbitrary scalar-tensor theories, including not only the (beyond) Horndeski theories known so far but also those that remain unexplored. The EFT is characterized by several time-dependent parameters, and each concrete theory corresponds to a specific choice of these parameters. Hence, observational constraints on EFT parameters can be recast into those on concrete theories. Such constraints have often been obtained from CMB, baryon acoustic oscillation~(BAO), and Type~Ia supernovae~(SNe~Ia) observations, assuming a specific cosmological background (e.g., $\Lambda$CDM or $w$CDM) and particular functional forms of the EFT parameters~\cite{Planck:2015bue,Planck:2018vyg}.

Along this line of thought, in this work we use the Hubble parameter~$H(z)$, as a function of redshift~$z$, inferred from cosmic chronometer~(CC) observations via the Gaussian process~(GP) regression technique to reconstruct the parameters of the EFT of dark energy associated with the background dynamics.\footnote{See Ref.~\cite{Durakovic:2019kqq} for a reconstruction of the EFT of inflation from cosmological data.} Our approach demonstrates how the relevant EFT parameters can be constrained directly from observational data with minimal theoretical assumptions and, unlike previous studies, without relying on specific cosmological models. This provides a new avenue for bridging observational cosmology and the EFT framework.

The rest of this paper is organized as follows.
In Sec.~\ref{sec:CC_with_GP}, we describe the basics of the CC observations and the GP technique.
In Sec.~\ref{sec:EFT_with_CC}, we discuss the reconstruction of the EFT functions, denoted by $\Lambda(z)$ and $c(z)$. As an example of a concrete scalar-tensor theory, we then focus on the quintessence model and demonstrate the reconstruction of the scalar-field potential.
Finally, we draw our conclusions in Sec.~\ref{sec:Conclusions}.

\section{\texorpdfstring{Reconstruction of $H(z)$ and $dH/dz$}{Reconstruction of H(z) and dH/dz}} \label{sec:CC_with_GP}

\subsection{Cosmic Chronometer} \label{subsec:CC}

Cosmic chronometer observations provide a direct measurement of the Hubble parameter~$H$ as a function of redshift~$z$, whose basic idea was introduced in Ref.~\cite{Jimenez:2001gg}. In a homogeneous and isotropic Universe described by the spatially flat FLRW metric,
    \begin{align}
    g_{\mu\nu}dx^\mu dx^\nu=-dt^2+a^2(t)d\bm{x}^2, \label{FLRW}
    \end{align}
the redshift is defined in terms of the scale factor~$a(t)$ as $1+z=1/a$. Note in passing that one could in principle include nonvanishing spatial curvature, but we do not do so here for simplicity. The Hubble parameter is then given by
\begin{align}
    \label{eq:Hubble}
    H(z)=\frac{1}{a}\frac{da}{dt}
    =-\frac{1}{1+z}\frac{dz}{dt}.
\end{align}
In the CC method, the redshift of each galaxy---and hence the redshift difference~$\Delta z$ between galaxies at closely spaced redshifts---is obtained directly from spectroscopy. In particular, one focuses on passively evolving galaxies with little to no star formation, so that their stellar populations can be treated as effectively single-aged. The age difference~$\Delta t$ between two such galaxies is then inferred by fitting their spectra with stellar population synthesis models. One subsequently estimates $dz/dt$ from the ratio~$\Delta z/\Delta t$, which determines the Hubble parameter~$H(z)$ at the corresponding redshift through Eq.~\eqref{eq:Hubble}.

\begin{figure}[htbp]
    \centering
    \includegraphics[width=1\linewidth]{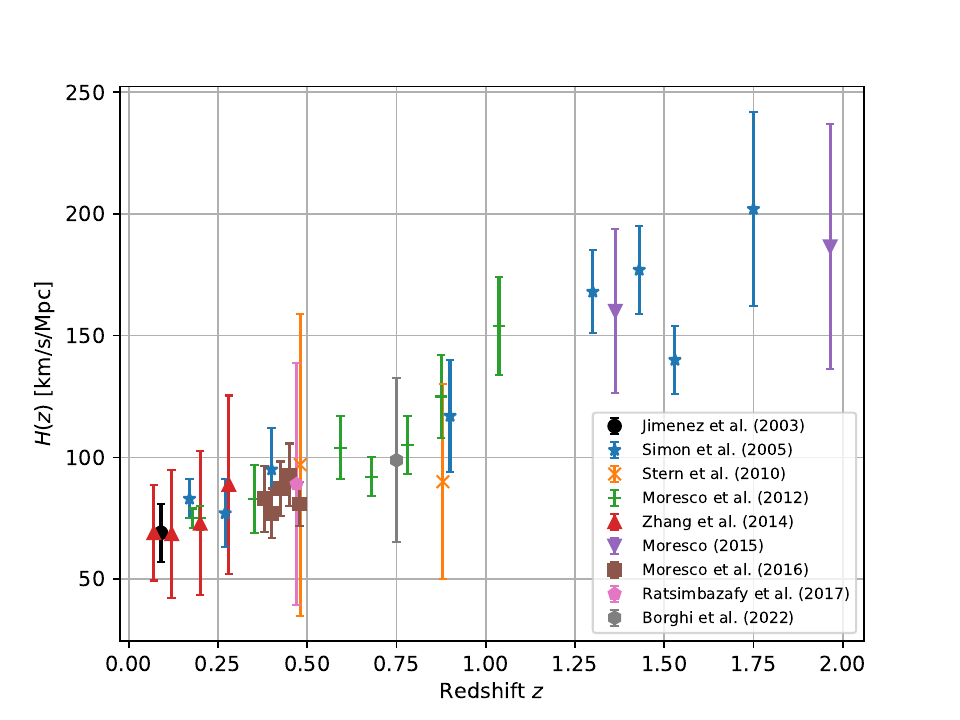}
    \caption{Summary of 32 CC data used in Ref.~\cite{Jalilvand:2022lfb}, taken from the following references, each indicated by a different label: Jimenez et al.~(2003)~\cite{Jimenez:2003iv}, Simon et al.~(2005)~\cite{Simon:2004tf}, Stern et al.~(2010)~\cite{Stern:2009ep}, Moresco et al.~(2012)~\cite{Moresco:2012jh}, Zhang et al.~(2014)~\cite{2014RAA....14.1221Z}, Moresco (2015)~\cite{Moresco:2015cya}, Moresco et al.~(2016)~\cite{Moresco:2016mzx}, Ratsimbazafy et al.~(2017)~\cite{Ratsimbazafy:2017vga}, and Borghi et al.~(2022)~\cite{Borghi:2021rft}.}
    \label{fig:CC_data}
\end{figure}

Although CC observations are subject to unavoidable systematic uncertainties---arising, e.g., from metallicity estimates, assumptions about the star formation history, and the stellar population synthesis model~\cite{Moresco:2020fbm}---they nevertheless offer several advantages:
\begin{enumerate}
    \item[(i)] They do not assume any cosmological model.
    \item[(ii)] They provide direct measurements of $H(z)$ over a range of redshifts, rather than at a single redshift.
    \item[(iii)] They allow for an independent determination of the Hubble constant~$H_0$ without relying on CMB data, thus breaking the degeneracy between the curvature parameter~$\Omega_K$ and $H_0$.
    \item[(iv)] They do not rely on any external cosmological calibration.
\end{enumerate}
These considerations motivate us to reconstruct $H(z)$ using the CC data, which we will discuss in detail in the next section. We adopt the same set of CC measurements as in Ref.~\cite{Jalilvand:2022lfb}, summarized in Fig.~\ref{fig:CC_data}. It is worth noting that, as more CC data become available in the future, the uncertainties in the reconstructed $H(z)$ will be further reduced \cite{Wang:2016iij}.

\subsection{Gaussian Process Regression} \label{subsec:GP}

In order to reconstruct the functional form of the Hubble parameter~$H(z)$ from the CC data, we apply GP regression, a nonparametric approach to inferring a function from discrete data points (see, e.g., Ref.~\cite{introduction_to_GP}). In the GP framework, the function values at any set of points are assumed to follow a multivariate Gaussian distribution characterized by a mean function, $m(z)$, which represents the expected value of the function at each redshift, and a covariance (kernel) function, $k(z,\tilde{z})$, which encodes correlations between function values at different points. Conditioning this prior on the given data yields the posterior mean and covariance, providing both a smooth reconstruction of $H(z)$ and its associated uncertainty. An additional advantage of GP regression is that, as long as the kernel function is differentiable, it is possible to obtain the error covariance matrix for the reconstructed function and its derivatives. In Refs.~\cite{Busti:2014dua, Wang:2016iij, Gomez-Valent:2018hwc,OColgain:2021pyh,Jalilvand:2022lfb}, this technique was used to estimate the Hubble constant~$H_0$. (See also Refs.~\cite{Shafieloo:2012ht, Liao:2019qoc, Pinho:2018unz, Mehrabi:2020zau} for applications of the GP technique to the reconstruction of other cosmological quantities.)

Let us introduce some commonly used kernel functions. One widely adopted choice is the squared exponential (SE) covariance function,
\begin{align}
    \label{eq:GP_SECF}
    k(z,\tilde{z})=\sigma_f^2\exp\left[-\frac{(z-\tilde{z})^2}{2l^2}\right],
\end{align}
where $z$ and $\tilde{z}$ are arbitrary redshifts within the reconstruction domain, and $\sigma_f$ and $l$ are hyperparameters optimized in the GP regression. Another example is the Mat{\'e}rn kernel~\cite{introduction_to_GP},
\begin{align}
    \label{eq:Matern}
    k(z,\tilde{z})=\sigma_f^2\frac{2^{1-\nu}}{\Gamma(\nu)}\left(\frac{\sqrt{2\nu}\abs{z-\tilde{z}}}{l}\right)^{\nu}K_{\nu}\left(\frac{\sqrt{2\nu}\abs{z-\tilde{z}}}{l}\right),
\end{align}
where $\Gamma$ and $K_{\nu}$ represent the gamma function and the modified Bessel function, respectively. The Mat{\'e}rn kernel involves a parameter~$\nu$, which controls the smoothness of the functions drawn from the GP. More concretely, $\nu$ is often chosen to be a half-integer, $\nu = \hat{\nu} + 1/2$ with an integer~$\hat{\nu}$, in which case the resulting functions are $\hat{\nu}$-times differentiable. In the limit~$\nu \to \infty$, the SE kernel~\eqref{eq:GP_SECF} is recovered. Note that although the kernel functions involve hyperparameters, the GP reconstruction remains nonparametric because no fixed functional form is imposed on the reconstructed function.

The choice of kernel and mean function is crucial in reconstructing observables using the GP. In Ref.~\cite{Busti:2014dua}, it was pointed out that the reconstructed $H(z)$ obtained from CC data can depend sensitively on the choice of kernel function. To reexamine this issue, we performed a GP reconstruction of $H(z)$ using the 32 CC data points shown in Fig.~\ref{fig:CC_data}. In particular, we tested the dependence on the kernel by considering the SE covariance function given in Eq.~\eqref{eq:GP_SECF} and several Mat{\'e}rn kernels with different values of~$\nu$. Figure~\ref{fig:H_dHdz_rec_multikernel_from_CC} shows the reconstructed $H(z)$ and its derivative~$dH/dz$ obtained with $m(z)=0$ and various kernel choices. We find that both $H(z)$ and $dH/dz$ are largely insensitive to the choice of kernel function, consistent with the result of Ref.~\cite{Jalilvand:2022lfb}. Nonetheless, as emphasized in Ref.~\cite{Shafieloo:2012ht}, the choice of kernel can introduce biases in the hyperparameter space due to the shape constraints it imposes.

\begin{figure}
    \centering
    \includegraphics[width=1.0\linewidth]{H_dHdz_reconstructing_multikernel_v2.pdf}
    \caption{Plots of $H(z)$ and $dH/dz$ reconstructed by GP with several kernel functions. Each colored band represents the $1\sigma$ uncertainty. We used $m(z)=0$ in this GP reconstruction. The black dots and error bars in the left panel represent the CC data shown in Fig.~\ref{fig:CC_data}.}
    \label{fig:H_dHdz_rec_multikernel_from_CC}
\end{figure}

\begin{figure}[htbp]
    \centering
    \includegraphics[width=1.0\linewidth]{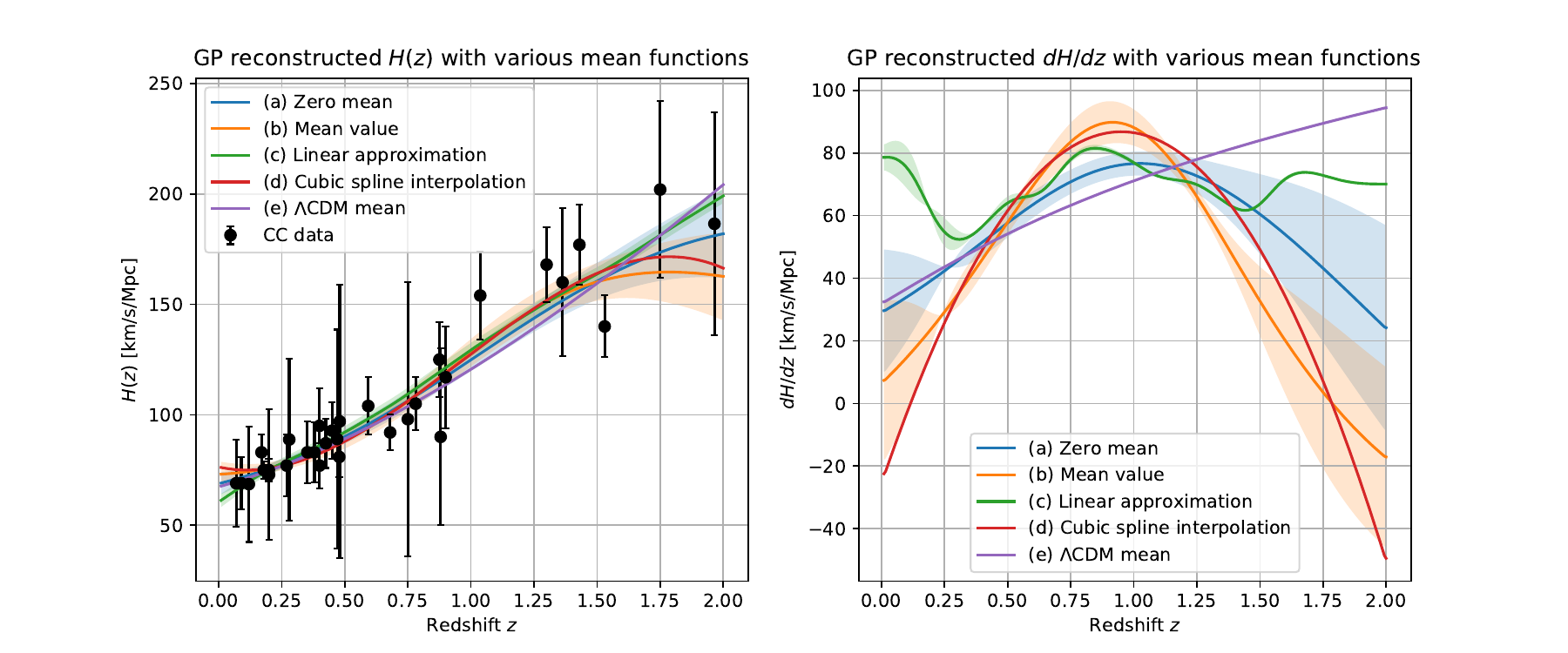}
    \caption{Plots of $H(z)$ and $dH/dz$ reconstructed by GP with several mean functions. Each colored band represents the $1\sigma$ uncertainty. We used the Mat{\'e}rn kernel with $\nu=3.5$ in this GP reconstruction. The black dots and error bars in the left panel represent the CC data shown in Fig.~\ref{fig:CC_data}.}
    \label{fig:H_dHdz_rec_multimean_from_CC}
\end{figure}

Similarly, the mean function can introduce biases or memory effects in the reconstructed results and therefore must be chosen with care~\cite{Shafieloo:2005nd,Shafieloo:2007cs,Shafieloo:2009hi,Shafieloo:2012ht,Hwang:2022hla}. To reexamine this issue, we reconstructed $H(z)$ and $dH/dz$ using the five mean functions listed in Table~\ref{tab:five_mean_functions}, adopting the Mat{\'e}rn kernel with $\nu = 3.5$ as the covariance function. Note in passing that option~(e), based on the $\Lambda$CDM model, was included tentatively for demonstration purposes; we omit it in the following analysis to avoid cosmological model dependence. As shown in Fig.~\ref{fig:H_dHdz_rec_multimean_from_CC}, the choice of mean function leads to noticeable biases in regions with sparse data coverage, particularly in the reconstruction of $dH/dz$.

\begin{table}[htbp]
    \centering
    \caption{Summary of the mean functions adopted in the GP reconstruction. For option~(e), we use $H_0=67.36\,{\rm km\,s^{-1}Mpc^{-1}}$ and $\Omega_{{\rm m}0}=0.315$ from the results of Planck Collaboration 2018~\cite{Planck:2018vyg}. In the analysis presented in Sec.~\ref{sec:EFT_with_CC} and thereafter, we omit option~(e), used here only as a benchmark, and adopt options~(a)--(d) to avoid cosmological model dependence.}
    \label{tab:five_mean_functions}
    \begin{tabular}{|c|l|}\hline
        Label & Mean function~$m(z)$ \\ \hline
        (a) & $m(z)=0$\\ \hline
        (b) & Mean value of the CC data\\ \hline
        (c) & Linear approximation of the CC data\\ \hline
        (d) & Cubic spline interpolation of the CC data\\ \hline
        (e) & $m(z)=H_0\sqrt{\Omega_{{\rm m}0}(1+z)^3+1-\Omega_{{\rm m}0}}$ (benchmark; this section only)\\ \hline
    \end{tabular}
\end{table}

\begin{figure}[htbp]
    \centering
    \includegraphics[width=1\linewidth]{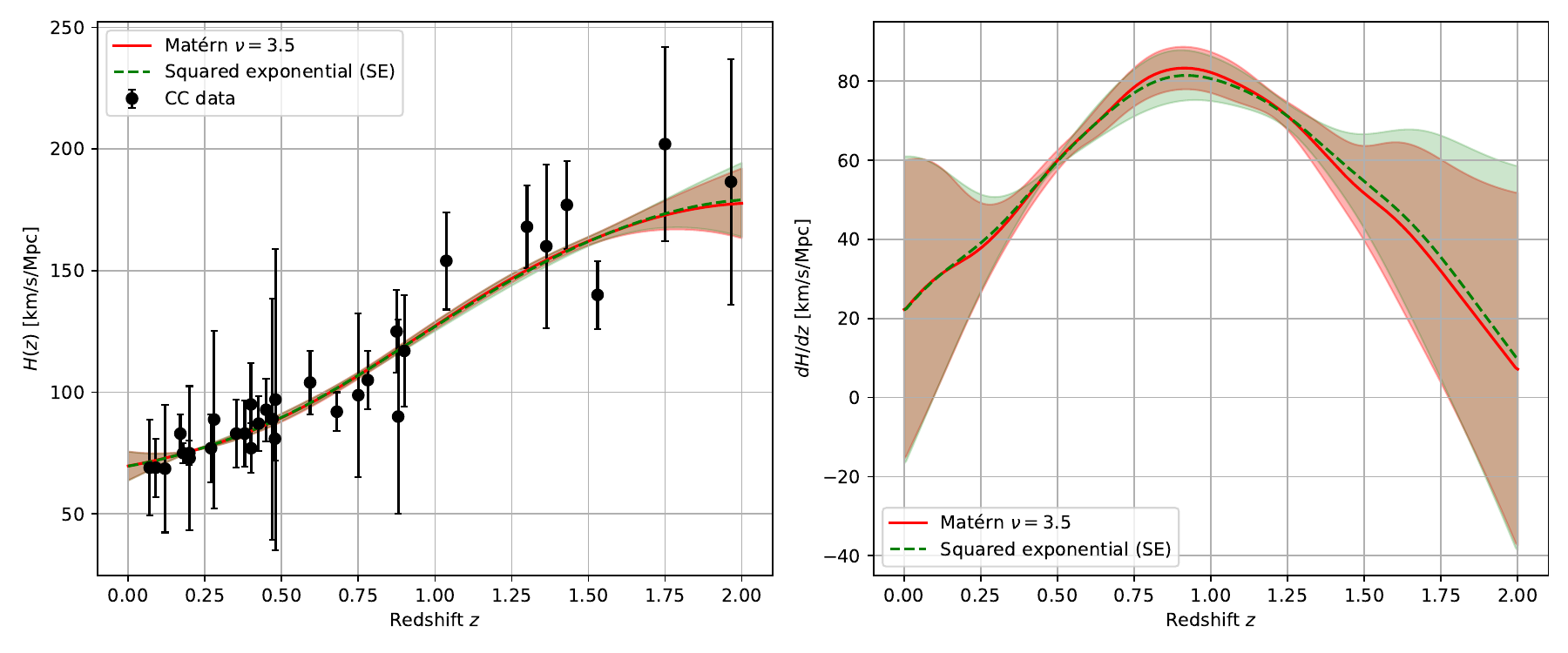}
    \caption{Plots of $H(z)$ and $dH/dz$ reconstructed by GP using the method of Ref.~\cite{Shafieloo:2012ht}. Each colored band represents the $1\sigma$ uncertainty. The black dots and error bars in the left panel represent the CC data shown in Fig.~\ref{fig:CC_data}.}
    \label{fig:shafieloomethod_H_dHdz}
\end{figure}

To mitigate the issues associated with the choice of kernel and mean function, we follow an iterative smoothing approach and a weighted averaging method based on the posterior distributions of hyperparameters, proposed in Ref.~\cite{Shafieloo:2012ht} (see also Refs.~\cite{Shafieloo:2005nd,Shafieloo:2007cs,Shafieloo:2009hi,Hwang:2022hla}). In this procedure, the mean function is iteratively smoothed over some scale in $\ln(1+z)$, starting from several initial guess functions, until the $\chi^2$ values converge within $\Delta\chi^2 = 2.3$, corresponding to the $1\sigma$ confidence level for two degrees of freedom. This process yields a data-driven mean function that is effectively free from memory of any specific cosmological model. The results are then averaged over the chosen initial guess functions and marginalized over the covariance hyperparameters~$(\sigma_f^2,l)$, using log-flat priors~$10^{-5} \le \sigma_f^2 \le 1$ and $10^{-2} \le l \le 10^{0.2}$. The resulting posterior-weighted reconstructions reduce biases associated with specific kernel or mean-function choices and provide accurate results with well-characterized uncertainties. The reconstructed $H(z)$ and $dH/dz$ are shown in Fig.~\ref{fig:shafieloomethod_H_dHdz}, where both the Mat{\'e}rn kernel with $\nu = 3.5$ and the SE covariance function were employed. For the initial guess mean functions, we used options~(a)--(d) in Table~\ref{tab:five_mean_functions}, omitting option~(e) as mentioned earlier. In the subsequent analysis, we adopt the Mat{\'e}rn kernel with $\nu = 3.5$, following Refs.~\cite{Shafieloo:2012ht,Hwang:2022hla}, since the SE covariance function tends to be overly smooth and may wash out genuine features of the data.

\section{Reconstruction of Background EFT Functions} \label{sec:EFT_with_CC}

\subsection{EFT of Dark Energy} \label{subsec:EFT}

The effective field theory of inflation/dark energy~\cite{Creminelli:2006xe,Cheung:2007st,Gubitosi:2012hu} is a model-independent framework for analyzing cosmological perturbations in scalar-tensor gravity.\footnote{The EFT has recently been extended to include vector-tensor theories~\cite{Aoki:2021wew,Aoki:2024ktc} or solids/fluids~\cite{Aoki:2022ipw}. The EFTs are distinguished by different symmetry breaking patterns.}
A key assumption is that the scalar field~$\phi$ has a timelike gradient, which allows us to choose the time coordinate so that a constant-$\phi$ hypersurface coincides with a constant-$t$ hypersurface. In this gauge, known as the unitary gauge, the residual symmetry is the spatial diffeomorphism invariance. Therefore, the gravitational action is constructed out of quantities that transform covariantly under the spatial diffeomorphisms, e.g., the extrinsic curvature~$K_{\mu\nu}$ and the three-dimensional Ricci tensor~${}^{(3)}\!R_{\mu\nu}$ associated with the constant-$t$ slicing. Written more explicitly, in terms of the unit timelike normal vector~$n_\mu\equiv -\partial_\mu t/\sqrt{-g^{00}}$ and the projection tensor~$h_{\mu\nu}\equiv g_{\mu\nu}+n_\mu n_\nu$, the extrinsic curvature is given by
\begin{align}
    \label{eq:Kij}
    K_{\mu\nu}=h^{\lambda}_{\mu}\nabla_{\lambda}n_{\nu},
\end{align}
where $\nabla_\mu$ denotes the covariant derivative associated with the spacetime metric~$g_{\mu\nu}$. Note in passing that the three-dimensional Riemann tensor is fully determined by ${}^{(3)}\!R_{\mu\nu}$ as the Weyl tensor identically vanishes in three-dimensions. Then, the EFT action in the unitary gauge can be schematically written as
    \begin{align}
    S=\int d^4x\sqrt{-g}\,{\cal L}(t,g^{00},K^\mu_\nu,{}^{(3)}\!R^\mu_\nu)
    +S_{\rm m}.
    \end{align}
Here, $S_{\rm m}$ represents the action of matter fields, which are assumed to be minimally coupled to gravity. Practically, one expands the action up to the necessary order in perturbations and derivatives. On a homogeneous and isotropic cosmological background described by the spatially flat FLRW metric~\eqref{FLRW}, the expansion coefficients are functions of $t$, which control the dynamics of perturbations at each order. Up to quadratic order in perturbations and leading order in derivatives, we have~\cite{Gubitosi:2012hu,Bloomfield:2012ff,Gleyzes:2013ooa,Tsujikawa:2014mba}
\begin{align}
    \label{eq:EFT_action}
    S=&\int d^4x\sqrt{-g}\left[\frac{M_{\ast}^2}{2}R-\Lambda-cg^{00}+\frac{M_2^4}{2}(\delta g^{00})^2-\frac{\bar{m}_1^3}{2}\delta K\delta g^{00}-\frac{\bar{M}_2^2}{2}\delta K^2\right.\nonumber\\
    &\left.-\frac{\bar{M}_3^2}{2}\delta K^{\mu}_{\nu}\delta K^{\nu}_{\mu}+\frac{\mu_1^2}{2}
    {}^{(3)}\!R\,\delta g^{00}+\frac{\bar{m}_5}{2}
    {}^{(3)}\!R\,\delta K+\cdots\right]
    +S_{\rm m},
\end{align}
where $K\equiv K^\mu_\mu$ denotes the trace of the extrinsic curvature, $\delta Q$ represents the perturbation of a quantity~$Q$, and the ellipsis indicates terms of higher order in perturbations and/or derivatives. It should be noted that the three-dimensional Ricci scalar~${}^{(3)}\!R$ vanishes at the background level, and is therefore purely perturbative. The coefficients~$M_*^2$, $\Lambda$, $c$, $M_2^4$, etc., are functions of $t$.\footnote{Note that parameters with even powers can, in principle, take either sign; the powers merely indicate mass dimensions. That said, we assume $M_*^2>0$ to ensure the linear stability of tensor perturbations.}

In what follows, we study the dynamics of the background spacetime based on the EFT. In this case, the only relevant EFT parameters are $M_*^2$, $\Lambda$, and $c$. We assume that the matter sector consists of a single barotropic perfect fluid with the energy density~$\rho_{\rm m}$ and the pressure~$p_{\rm m}$. The background equations of motion (i.e., the tadpole cancellation conditions) are given by
\begin{align}
    \label{eq:Friedmann_EFT}
    3M_{\ast}^2H^2&=\rho_{\rm m}+\Lambda+c-3H\frac{d{M}_{\ast}^2}{dt},\\
    \label{eq:dotFriedmann_EFT}
    \left(2\frac{dH}{dt}+3H^2\right)M_{\ast}^2&=-p_{\rm m}+\Lambda-c-2H\frac{d{M}_{\ast}^2}{dt}-\frac{d^2{M}_{\ast}^2}{dt^2}.
\end{align}
Recall that we restrict ourselves to the spatially flat FLRW metric~\eqref{FLRW}, and therefore terms involving the curvature parameter~$\Omega_K$ are not included. Also, provided that the fluid is minimally coupled to gravity, we have the following continuity equation:
\begin{align}
    \label{eq:conserva_matter}
    \frac{d\rho_{\rm m}}{dt}+3H(\rho_{\rm m}+p_{\rm m})=0.
\end{align}

With the setup described above, let us proceed to reconstruct the EFT functions using the Hubble parameter~$H(z)$ obtained from the CC data. It should be noted that, since only two independent equations, Eqs.~\eqref{eq:Friedmann_EFT} and \eqref{eq:dotFriedmann_EFT}, are available on an FLRW background, the three EFT functions~$M_*^2$, $\Lambda$, and $c$ cannot all be determined simultaneously from the background dynamics. In practice, we specify the functional form of $M_*^2$, as it directly controls the strength of gravity and its possible time variation is sometimes considered in connection with observational constraints. The functions~$\Lambda$ and $c$ are then consistently expressed in terms of $H(z)$ and its derivatives as follows:
\begin{align}
    \label{eq:Lambda_z}
    \Lambda(z)&=-\frac{1}{2}(\rho_{\rm m}-p_{\rm m})+HM_{\ast}^2\left[3H-(1+z)\frac{dH}{dz}\right]\nonumber\\
    &\hspace{1cm}-2H^2(1+z)\frac{dM_{\ast}^2}{dz}+\frac{1}{2}H(1+z)^2\frac{dH}{dz}\frac{dM_{\ast}^2}{dz}+\frac{1}{2}H^2(1+z)^2\frac{d^2M_{\ast}^2}{dz^2},\\
    \label{eq:c_z}
    c(z)&=-\frac{1}{2}(\rho_{\rm m}+p_{\rm m})+H(1+z)\frac{dH}{dz}M_{\ast}^2\nonumber\\
    &\hspace{1cm}-H^2(1+z)\frac{dM_{\ast}^2}{dz}-\frac{1}{2}H(1+z)^2\frac{dH}{dz}\frac{dM_{\ast}^2}{dz}-\frac{1}{2}H^2(1+z)^2\frac{d^2M_{\ast}^2}{dz^2},
\end{align}
where functions of $t$ have been recast as functions of $z$ through $a(t)=1/(1+z)$, and we have used the relation~$\frac{d}{dt}=-H(z)(1+z)\frac{d}{dz}$.
These formulae allow us to evaluate the uncertainties of $\Lambda(z)$ and $c(z)$ using the predictive mean and full covariance matrix obtained from the CC reconstruction of $H(z)$. Specifically, we generated Monte Carlo samples from the Gaussian distribution defined by this mean and covariance using sufficiently fine redshift bins, computed $dH/dz$ for each realization using a finite-difference approximation, and then derived the corresponding $68\%$ and $95\%$ confidence intervals for $\Lambda(z)$ and $c(z)$. The numbers of samples and redshift bins were chosen to be sufficiently large to ensure that the results are stable under variations of these choices.

As can be seen from Eqs.~\eqref{eq:Lambda_z} and \eqref{eq:c_z}, the matter field must be specified to reconstruct the EFT parameters. In what follows, we consider a pressureless dust as the matter component, i.e., $p_{\rm m}(z)=0$ and $\rho_{\rm m}(z)=\rho_{{\rm m}0}(1+z)^3=3M_{\rm Pl}^2H_0^2\Omega_{{\rm m}0}(1+z)^3$, where $\Omega_{{\rm m}0}$ denotes the present matter density parameter. Since $\Omega_{{\rm m}0}$ is not specified a priori, we adopt three representative values, $\Omega_{{\rm m}0}\in\{0.25,0.30,0.35\}$.\footnote{Instead of $\Omega_{{\rm m}0}$ itself, one could choose $\Omega_{{\rm m}0}h^2$ as an independent parameter, with $h$ being the Hubble parameter normalized by $100\,{\rm km\,s^{-1}\,Mpc^{-1}}$. As far as we have investigated, no significant difference was found between the two choices.}
In particular, the choice~$\Omega_{{\rm m}0}=0.30$ is close to the value reported by the Planck Collaboration ($\Omega_{{\rm m}0}=0.315 \pm 0.007$)~\cite{Planck:2018vyg}.

\subsection{Reconstruction under Minimal Coupling} \label{subsec:Mast_Mpl}

\begin{figure}[htbp]
    \centering
    \includegraphics[width=1.0\linewidth]{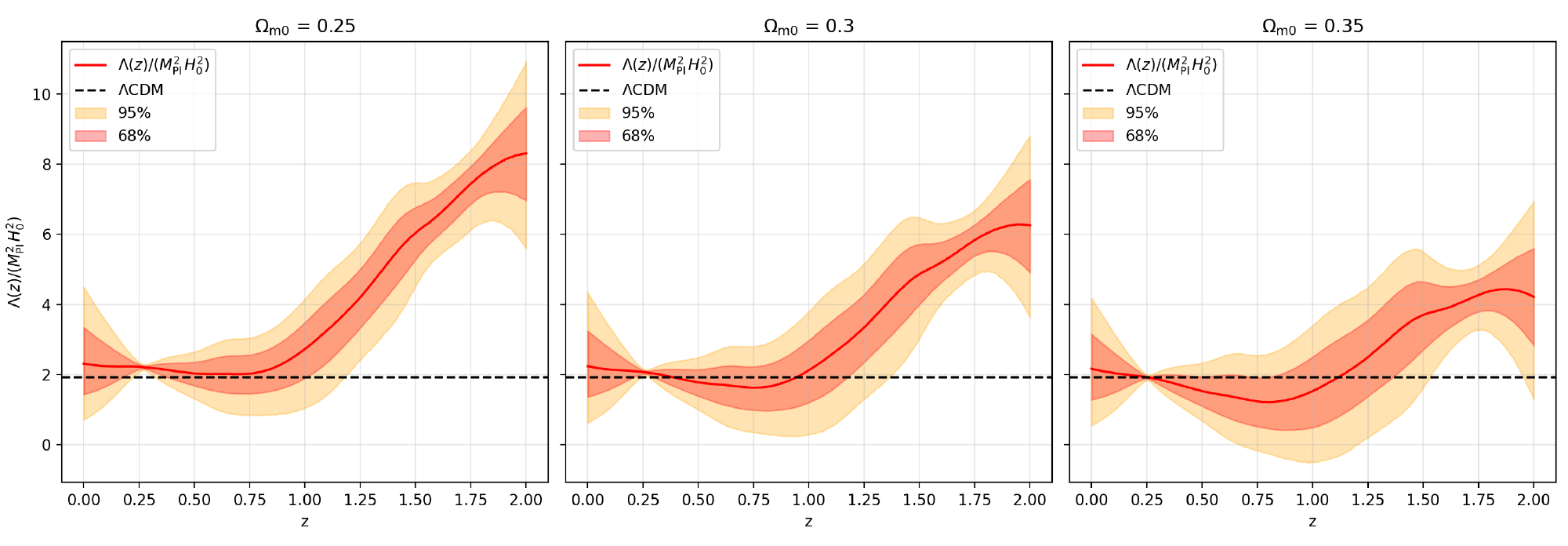}
    \caption{Plots of $\Lambda(z)/(M_{\rm Pl}^2H_0^2)$ reconstructed from CC data using Eq.~\eqref{eq:Lambda_z_simple} for $\Omega_{{\rm m}0} \in \{0.25,0.30,0.35\}$. The black dashed line represents the constant value in the $\Lambda$CDM model, $3(1-\Omega_{{\rm m}0})$, evaluated at the central Planck 2018 value~$\Omega_{{\rm m}0}=0.315\,(\pm 0.007)$~\cite{Planck:2018vyg}.}
    \label{fig:Lambda_z_simple_case}
\end{figure}

\begin{figure}[htbp]
    \centering
    \includegraphics[width=1.0\linewidth]{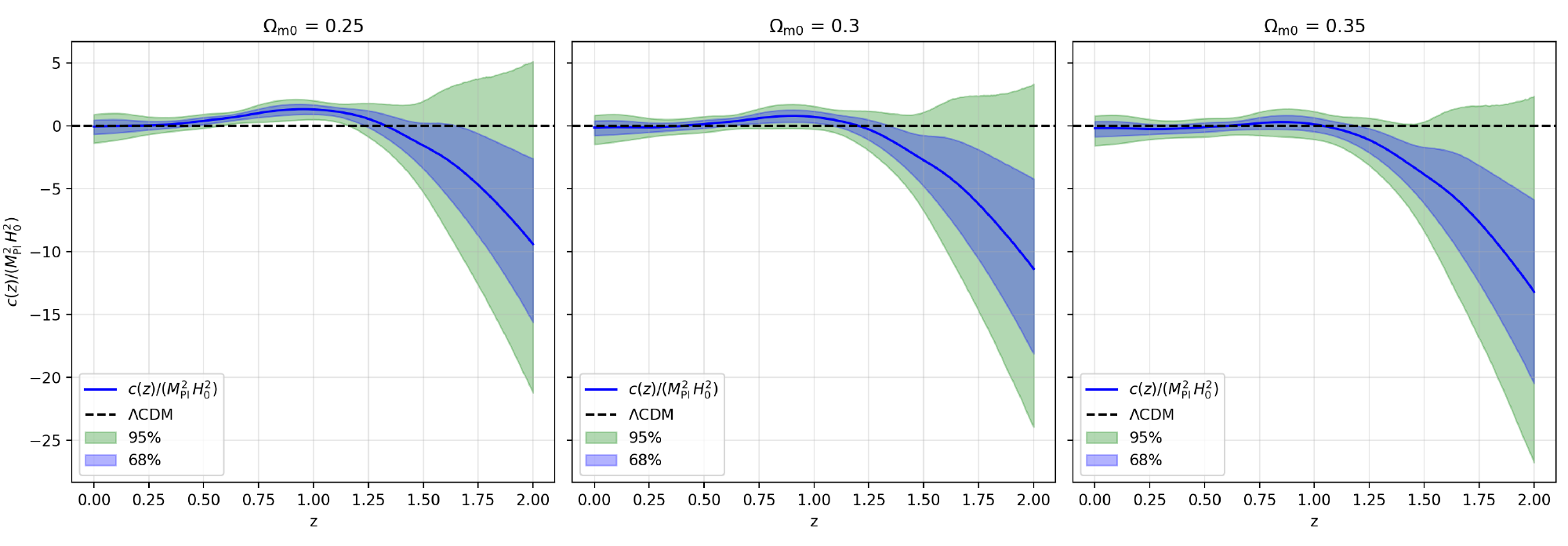}
    \caption{Plots of $c(z)/(M_{\rm Pl}^2H_0^2)$ reconstructed from CC data using Eq.~\eqref{eq:c_z_simple} for $\Omega_{{\rm m}0} \in \{0.25,0.30,0.35\}$. The black dashed line represents the $\Lambda$CDM prediction, i.e., $c(z)=0$.}
    \label{fig:c_z_simple_case}
\end{figure}

For demonstration purposes, let us assume $M_{\ast}^2(z) = M_{\rm Pl}^2 = \mathrm{const}$, with $M_{\rm Pl}$ denoting the reduced Planck mass, corresponding to minimal coupling to gravity (see the \hyperref[app:Mast_power_law]{Appendix} for an analysis with time-dependent $M_*^2$). In this case, Eqs.~\eqref{eq:Lambda_z} and \eqref{eq:c_z} are simplified as
\begin{align}
    \label{eq:Lambda_z_simple}
    \Lambda(z)&=-\frac{3}{2}M_{\rm Pl}^2H_0^2\Omega_{{\rm m}0}(1+z)^3+M_{\rm Pl}^2H\left[3H-(1+z)\frac{dH}{dz}\right],\\
    \label{eq:c_z_simple}
    c(z)&=-\frac{3}{2}M_{\rm Pl}^2H_0^2\Omega_{{\rm m}0}(1+z)^3+M_{\rm Pl}^2H(1+z)\frac{dH}{dz}.
\end{align}
Figures~\ref{fig:Lambda_z_simple_case} and \ref{fig:c_z_simple_case} show $\Lambda(z)$ and $c(z)$ normalized by $M_{\rm Pl}^2H_0^2$ reconstructed from CC data, using the mean value~$H_0=69.7\pm5.93~{\rm km/s/Mpc}$ estimated from the CC data by GP as a reference value.
From Fig.~\ref{fig:Lambda_z_simple_case}, we find that the $\Lambda(z)/(M_{\rm Pl}^2H_0^2)$ inferred from the CC data agrees within $2\sigma$ with the constant value obtained from the $\Lambda$CDM model using the Planck 2018 parameter ($\Omega_{{\rm m}0} = 0.315$)~\cite{Planck:2018vyg} at lower redshifts ($z \lesssim 1.25$). This indicates that the CC data alone do not provide compelling evidence for a dynamical dark energy component, in contrast to the trend suggested by the Dark Energy Spectroscopic Instrument (DESI) BAO measurements~\cite{DESI:2024lzq, DESI:2024mwx}.
Note that the uncertainty band for $\Lambda(z)$ is locally very narrow around $z\simeq 0.25$ due to the strong correlation between $H$ and $dH/dz$. At higher redshifts ($z \gtrsim 1.25$), $\Lambda(z)$ deviates from the $\Lambda$CDM model by more than $2\sigma$, which is likely driven by the sparse CC data in that redshift range.
Figure~\ref{fig:c_z_simple_case} shows that $c(z)/(M_{\rm Pl}^2H_0^2)$ is consistent with zero within $2\sigma$, in agreement with the $\Lambda$CDM prediction.

At higher redshifts ($z \gtrsim 1.25$), the dependence on $\Omega_{\mathrm{m}0}$ and the apparent deviations of $c(z)$ from the $\Lambda$CDM prediction become more pronounced. This tendency, however, may again be attributed to the limited sampling of CC data in that redshift range. Therefore, additional observational data are essential for achieving more reliable reconstructions at high redshifts. At lower redshifts ($z \lesssim 1.25$), the reconstructed $\Lambda(z)$ and $c(z)$ remain statistically consistent across different assumptions of $\Omega_{\mathrm{m}0}$. This robustness is supported by the abundance of CC data and the stability of the GP reconstructions in this regime. Consequently, conclusions based on the EFT functions in this redshift range are not significantly affected by the assumed values of $\Omega_{\mathrm{m}0}$.

\subsection{Application: Reconstruction of the Quintessence Model} \label{subsec:Reconstruction_phi_V}

Let us now apply the EFT functions~$\Lambda(z)$ and $c(z)$ obtained in the previous subsection to reconstruct the quintessence model,\footnote{See also Refs.~\cite{Starobinsky:1998fr,Nakamura:1998mt,Chiba:2000im} for earlier work on the reconstruction of the scalar-field potential in the quintessence model.}
whose Lagrangian is given by~\cite{Fujii:1982ms, Ratra:1987rm, Chiba:1997ej, Copeland:1997et, Ferreira:1997hj, Caldwell:1997ii, Zlatev:1998tr}
    \begin{align}
    {\cal L}=-\frac{1}{2}g^{\mu\nu}\partial_\mu\phi\,\partial_\nu\phi-V(\phi),
    \end{align}
which is represented in the EFT by
    \begin{align}
    \Lambda=V(\phi(t)), \qquad
    c=\frac{1}{2}\left(\frac{d\phi}{dt}\right)^2. \label{dictionary_quintessence}
    \end{align}
Since $\Lambda$ and $c$ are known as functions of the redshift~$z$, we can infer $\phi$ as a function of $z$, and subsequently reconstruct $V$ as a function of $\phi$. Written explicitly, assuming $d\phi/dt > 0$,
    \begin{align}
    \phi(z)-\phi_0&=-\int_{0}^{z}\frac{\sqrt{2c(z')}}{(1+z')H(z')}dz', \label{phi_c} \\
    V(\phi)&=\Lambda(z), \label{V_Lambda}
    \end{align}
where $\phi_0$ denotes the value of $\phi$ at $z=0$. Here, in the second equation, the redshift $z$ is understood as a function of $\phi$ through the first equation. Note in passing that $\phi_0$ is arbitrary, and the reconstruction is defined only up to this constant offset. It should also be noted that the domain of the reconstructed potential is limited to the field range covered by $\phi(z)$ obtained from Eq.~\eqref{phi_c}.

It is worth emphasizing that this reconstruction should be interpreted as an illustrative application of the EFT framework rather than a stringent constraint on quintessence itself. In fact, the behavior of the reconstructed EFT functions already suggests that CC data alone favor an evolution close to that of a cosmological constant: as shown in Fig.~\ref{fig:c_z_simple_case}, the function~$c(z)$ fluctuates around zero within its uncertainty, whereas Eq.~\eqref{dictionary_quintessence} requires $c$ to remain positive in the quintessence model. This indicates an almost frozen scalar field and makes a direct reconstruction of the quintessence dynamics problematic with current CC data. To address this difficulty, we fix $\Omega_{{\rm m}0}=0.30$ for concreteness and use the upper boundary of the 1$\sigma$ confidence interval of $c(z)$, rather than the central value, as a proxy for the parameter~$c$ used in the reconstruction procedure. With this prescription, the reconstruction remains well-defined for $z \in [0, 1.33]$.

\begin{figure}[htbp]
    \centering
    \includegraphics[width=1\linewidth]{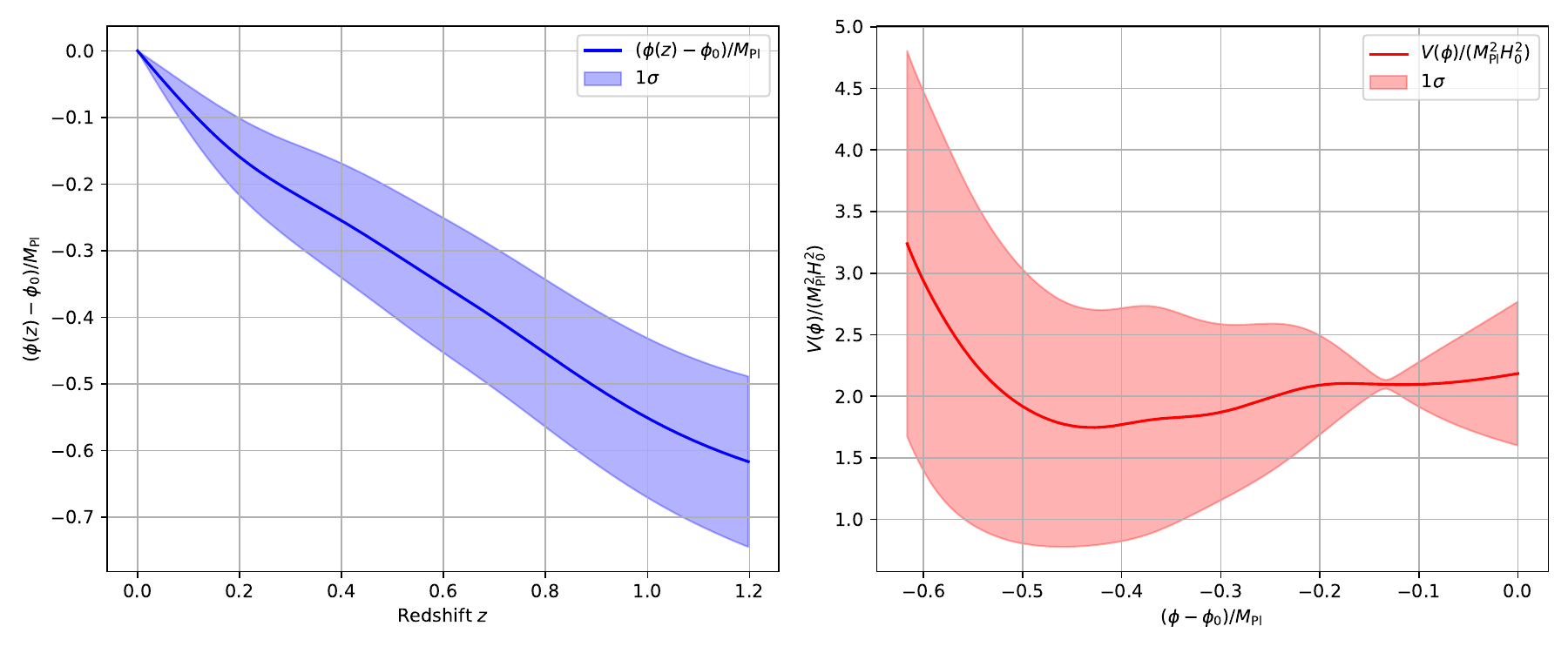}
    \caption{
    Plots of the reconstructed $\phi(z)$ and $V(\phi)$ for $\Omega_{{\rm m}0}=0.30$. In the reconstruction, the upper boundary of the 1$\sigma$ confidence interval of $c(z)$ is used as a proxy for the parameter~$c$. The shaded regions denote the 1$\sigma$ uncertainties.}
    \label{fig:Quitessence_phi_V_Om03}
\end{figure}

The reconstructed scalar field and potential are shown in Fig.~\ref{fig:Quitessence_phi_V_Om03}. The evolution of $\phi(z)$ is mild, with the total change in $\phi$ (in units of $M_{\rm Pl}$) remaining at ${\cal O}(10^{-1})$ over the plotted range of $z$. This weak evolution reflects the fact that the EFT function~$c(z)$ inferred from the CC data is close to zero within its uncertainty, indicating an almost frozen scalar field. Consequently, the reconstructed potential~$V(\phi)$ appears nearly flat, and the apparent structure in the curve is largely driven by the uncertainty in the mapping from $z$ to $\phi$. Although the errors in $\phi(z)$ and $V(\phi)$ remain sizable, future CC measurements with improved redshift coverage and precision are expected to reduce these uncertainties and enable more robust reconstructions. Overall, the present results suggest that current CC data are consistent with a nearly constant dark-energy component and do not favor appreciable scalar-field dynamics.

\section{Conclusions} \label{sec:Conclusions}

The EFT of dark energy provides a universal framework for describing cosmological dynamics in scalar-tensor gravity. In this work, we have proposed a method to reconstruct the EFT parameters using CC measurements. Specifically, by employing the Hubble parameter~$H(z)$ and its derivative~$dH/dz$ inferred from CC data through the Gaussian process regression technique, we have reconstructed the EFT functions that govern the background evolution. Our approach does not rely on any specific cosmological model such as $\Lambda$CDM or $w$CDM, in sharp contrast to previous studies~\cite{Raveri:2014cka, Planck:2015bue, Planck:2018vyg, Hiramatsu:2022fgn}, where the EFT functions were derived from other observations (e.g., CMB and BAO) under assumptions about specific functional forms for these parameters and the background evolution.

To illustrate our approach in detail, we have applied it to the EFT action~\eqref{eq:EFT_action}, which involves three parameters that determine the background dynamics, namely $M_*^2(z)$, $\Lambda(z)$, and $c(z)$. In the main text, we have set $M_*^2$ equal to the (reduced) Planck mass squared, $M_{\rm Pl}^2$, corresponding to a scalar field minimally coupled to gravity, and reconstructed the remaining parameters~$\Lambda$ and $c$ as functions of redshift~$z$ using Eqs.~\eqref{eq:Lambda_z_simple} and~\eqref{eq:c_z_simple}. In doing so, a pressureless dust component was included, adopting three representative values of the matter density parameter, $\Omega_{{\rm m}0}\in\{0.25,0.30,0.35\}$. (See the \hyperref[app:Mast_power_law]{Appendix} for an analysis with time-dependent $M_*^2$.) As shown in Figs.~\ref{fig:Lambda_z_simple_case} and~\ref{fig:c_z_simple_case}, the reconstructed parameters~$\Lambda(z)$ and $c(z)$ are consistent with those in the $\Lambda$CDM model within $2\sigma$. At higher redshifts ($z \gtrsim 1.25$), however, the reconstructed $\Lambda(z)$ and $c(z)$ exhibit a notable dependence on the assumed values of $\Omega_{{\rm m}0}$. That said, this tendency should not be overinterpreted, as it may simply reflect the sparsity of the CC data in this redshift range. Conversely, at lower redshifts ($z \lesssim 1.25$), the reconstructions of $\Lambda(z)$ and $c(z)$ are largely insensitive to variations in these parameters, providing a more reliable basis for physical interpretation. In Sec.~\ref{subsec:Reconstruction_phi_V}, we have used the quintessence model as a representative example to reconstruct the functional forms of the scalar field~$\phi(z)$ and its potential~$V(\phi)$ from the reconstructed $\Lambda(z)$ and $c(z)$, without specifying the functional form of $V(\phi)$. Our approach can also be applied to other concrete models, such as Horndeski theories, providing a powerful means to constrain modified gravity directly from observational data.

In summary, we have established a general, model-independent framework for testing the accelerated expansion of the Universe at the background level. Our results also highlight the importance of measuring the Hubble parameter~$H(z)$ in a model-independent manner, for instance through CC measurements. A current limitation of our approach is that the CC data are subject to relatively large observational uncertainties, which in turn lead to significant uncertainties in the reconstructed $\Lambda(z)$ and $c(z)$. However, as noted in Ref.~\cite{Wang:2016iij}, improving the quantity and quality of observational data would enhance the accuracy of the reconstructed $H(z)$, and a corresponding improvement can be expected for $\Lambda(z)$ and $c(z)$ as well. As a direction for future work, it would be intriguing to investigate how the uncertainties in the reconstructions of the EFT functions~$\Lambda(z)$ and $c(z)$ can be reduced by increasing the number of CC measurements. Moreover, although specifying cosmological models may become necessary, it will be particularly interesting to combine CC data with other observations, such as BAO (e.g., from DESI, whose recent results hint at dynamical dark energy~\cite{DESI:2024lzq, DESI:2024mwx}) and SNe~Ia, to achieve a more comprehensive reconstruction of the EFT functions.

\section*{Acknowledgments}
We thank Takeshi Chiba and Takashi Hiramatsu for insightful comments and discussions.
F.O.~was supported by individual research funding from Nihon University.
K.T.~was supported in part by Japan Society for the Promotion of Science KAKENHI Grant No.\ JP23K13101.


\appendix
\renewcommand{\theequation}{A.\arabic{equation}}

\section*{Appendix: Reconstruction under Non-minimal Coupling} \label{app:Mast_power_law}

In Sec.~\ref{subsec:Mast_Mpl}, we reconstructed the EFT functions~$\Lambda(z)$ and $c(z)$ under the assumption that $M_{\ast}^2(z)$ is constant, corresponding to minimal coupling to gravity. In this appendix, we examine how a time-dependent $M_{\ast}^2(z)$ affects the reconstruction. As an illustrative example, we adopt the following parameterization~\cite{Planck:2015bue,Planck:2018vyg}:\footnote{A brief comment on this parametrization is in order. In the EFT of cosmological perturbations described by the action~\eqref{eq:EFT_action}, one often defines $M^2\equiv M_*^2 - M_3^2$, which corresponds to the effective Planck mass squared for tensor perturbations. The time evolution of $M^2$ is characterized by the dimensionless parameter~$\alpha_M \equiv (H M^2)^{-1} dM^2/dt$. When applying the EFT to late-time cosmology, one typically sets $M_3^2=0$ so that gravitational waves propagate at the speed of light, yielding $M^2 = M_*^2$. This assumption is strongly supported by the multimessenger observation of GW170817 and its electromagnetic counterpart, which constrained the tensor propagation speed to coincide with the speed of light to within parts in $10^{15}$~\cite{LIGOScientific:2017vwq,LIGOScientific:2017ync,LIGOScientific:2017zic}. For the parameterization in Eq.~\eqref{eq:Mast}, we obtain $\alpha_M = \alpha_{M0}(1+z)^{-s}$; thus $\alpha_{M0}$ directly corresponds to the present value of $\alpha_M$. The limit~$s\to 0$ recovers the case of constant $\alpha_M$.}
\begin{align}
    \label{eq:Mast}
    M_{\ast}^2(z)=M_{\rm Pl}^2\exp\left[\frac{\alpha_{M0}}{s}\left((1+z)^{-s}-1\right)\right],
\end{align}
where $\alpha_{M0}$ and $s$ are free parameters, and we follow Refs.~\cite{Planck:2015bue,Planck:2018vyg} in assuming $s>0$. The minimal-coupling case~$M_{\ast}^2(z)=M_{\rm Pl}^2$ is recovered by setting $\alpha_{M0}=0$. Substituting Eq.~\eqref{eq:Mast} into Eqs.~\eqref{eq:Lambda_z} and~\eqref{eq:c_z} yields the corresponding expressions for $\Lambda(z)$ and $c(z)$.

\begin{figure}[htbp]
    \centering
    \includegraphics[width=1.0\linewidth]{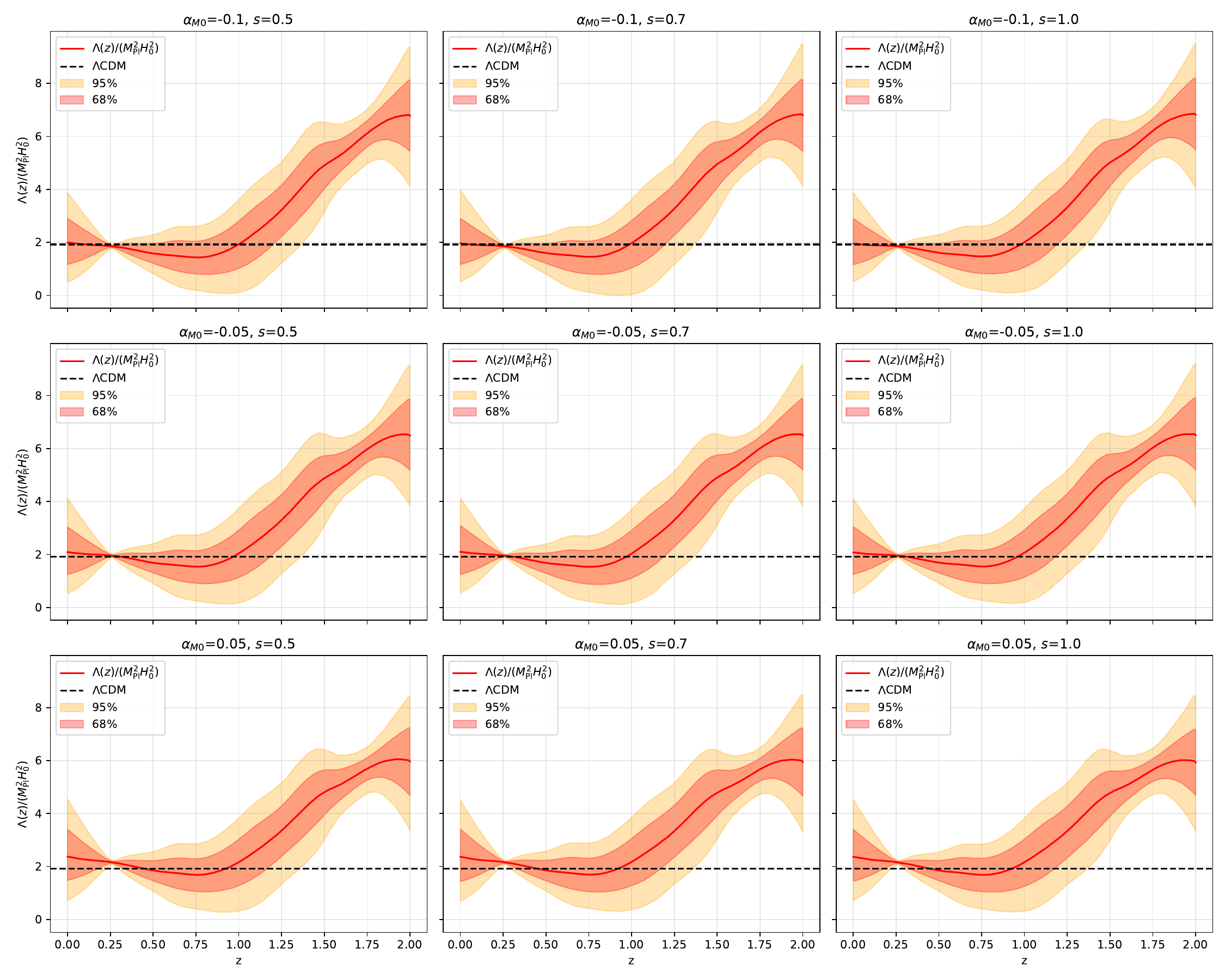}
    \caption{Plots of the reconstructed $\Lambda(z)$ using Eq.~\eqref{eq:Mast} for $\Omega_{{\rm m}0}=0.30$. Each panel corresponds to a different choice of $(\alpha_{M0}, s)$: from left to right, $\alpha_{M0} = -0.1, -0.05, 0.1$, and from top to bottom, $s = 0.5, 0.7, 1.0$, consistent with the Planck 2018 constraints~\cite{Planck:2018vyg}.}
    \label{fig:Lambda_Mastcase_Om03_threealphaM0_threes}
\end{figure}

\begin{figure}[htbp]
    \centering
    \includegraphics[width=1\linewidth]{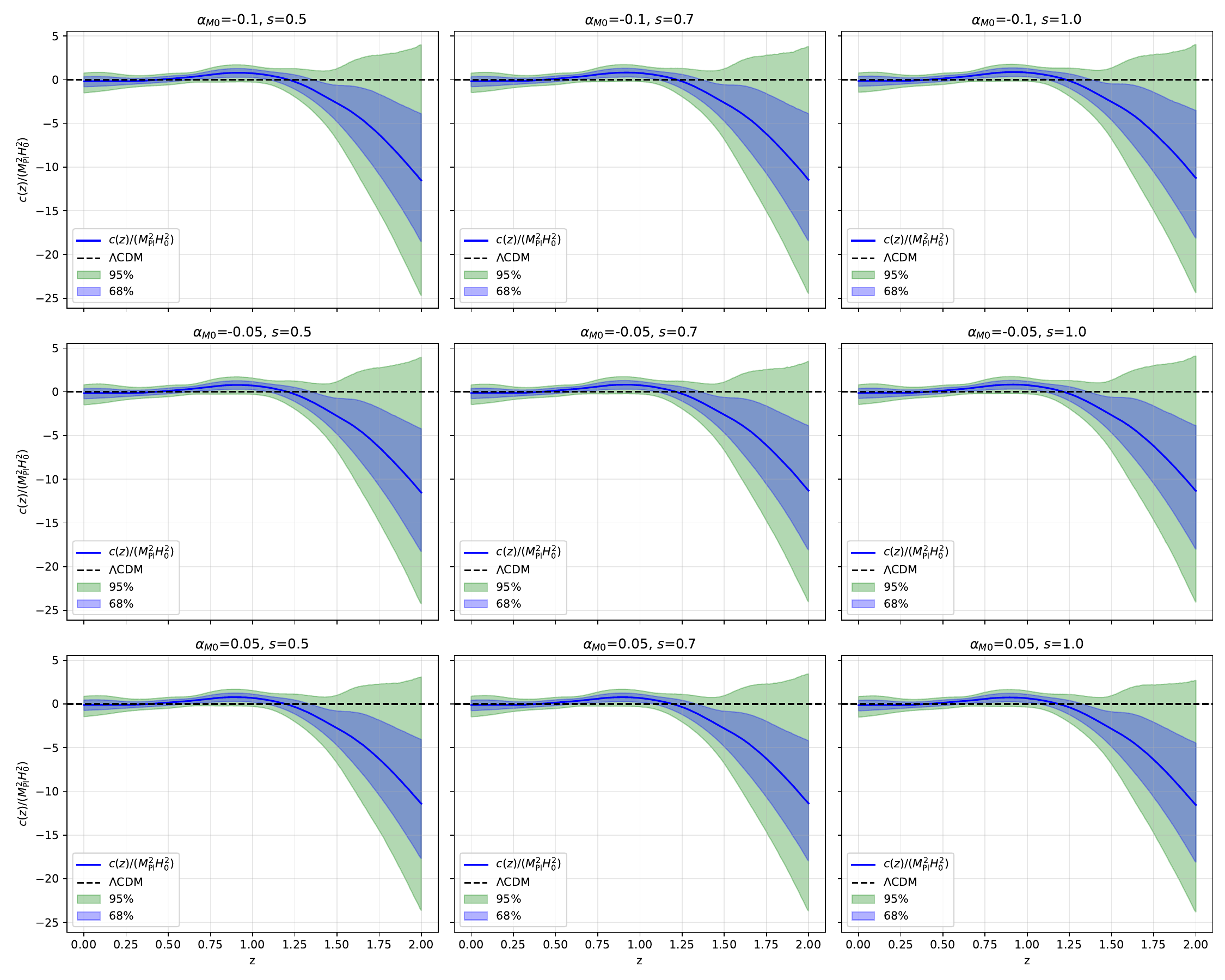}
    \caption{Plots of the reconstructed $c(z)$ using Eq.~\eqref{eq:Mast} for $\Omega_{{\rm m}0}=0.30$. Each panel corresponds to a different choice of $(\alpha_{M0}, s)$: from left to right, $\alpha_{M0} = -0.1, -0.05, 0.1$, and from top to bottom, $s = 0.5, 0.7, 1.0$, consistent with the Planck 2018 constraints~\cite{Planck:2018vyg}.}
    \label{fig:cz_Mastcase_Om03_threealphaM0_threes}
\end{figure}

In Figs.~\ref{fig:Lambda_Mastcase_Om03_threealphaM0_threes} and \ref{fig:cz_Mastcase_Om03_threealphaM0_threes}, we show the reconstructed $\Lambda(z)$ and $c(z)$ for $\Omega_{{\rm m}0}=0.30$ when the time dependence of $M_{\ast}^2(z)$ given in Eq.~\eqref{eq:Mast} is taken into account. We consider $\alpha_{M0}\in\{-0.1,-0.05,0.1\}$ and $s\in\{0.5,0.7,1.0\}$, choices consistent with the Planck 2018 constraints~\cite{Planck:2018vyg}. For these values of $\alpha_{M0}$ and $s$, the reconstructed $\Lambda(z)$ and $c(z)$ remain within $2\sigma$ of the $\Lambda$CDM prediction. Similarly to the minimal-coupling case studied in Sec.~\ref{subsec:Mast_Mpl}, the reconstructed $\Lambda(z)$ and $c(z)$ exhibit large uncertainties at higher redshifts ($z \gtrsim 1.25$), owing to the sparse sampling of CC data in that regime. Nonetheless, the reconstruction remains relatively insensitive to variations in $\alpha_{M0}$ and $s$, indicating that the EFT functions are robust within this redshift range.

\bibliography{refs}

\end{document}